\documentclass[aps,floatfix,nofootinbib,twocolumn]{revtex4}

\usepackage{graphicx}
\usepackage{epic}
\usepackage{eepic}
\usepackage{latexsym}

\newcommand{\eq}[1]{(\ref{#1})}
\newcommand{\be}{\begin{equation}}
\newcommand{\ee}{\end{equation}}
\newcommand{\bea}{\begin{eqnarray}}
\newcommand{\eea}{\end{eqnarray}}

\newcommand{\vs}[1]{\vspace{#1 mm}}
\newcommand{\hs}[1]{\hspace{#1 mm}}

\def\a{\alpha}

\def\D{\Delta}

\def\fr{\frac}

\def\l{\lambda}

\def\m{\mu}
\def\n{\nu}

\def\s{\sigma}
\def\S{\Sigma}
\def\t{\tau}
\def\th{\theta}

\def\O{\Omega}

\def\del{\partial}

\let\bm=\bibitem
\def\nn{\nonumber}

\begin{document}

\title{Boundary Effects in Local Inflation\\
 and Spectrum of Density Perturbations}

\author{Erol Ertan$^{1}$}
\email[]{ertan.erol@gmail.com}
\author{Ali Kaya$^{1,2}$\vs{3}}
\email[]{ali.kaya@boun.edu.tr}
\affiliation{$^{1}$Bo\~{g}azi\c{c}i University, Department of Physics, \\ 34342,
Bebek, \.Istanbul, Turkey\vs{3}\\$^{2}$Feza G\"{u}rsey Institute,\\
Emek Mah. No:68, \c{C}engelk\"{o}y, \.Istanbul, Turkey\vs{3}}

\date{\today}

\begin{abstract}

We observe that when a local patch in a radiation filled
Robertson-Walker universe inflates by some reason, outside
perturbations can enter into the inflating region. Generally, the
physical wavelengths of these perturbations become larger than the
Hubble radius  as they cross into the inflating space and their
amplitudes freeze out immediately. It turns out that the
corresponding power spectrum is not scale invariant. Although these
perturbations cannot  reach out to a distance inner observer
shielded by a de Sitter horizon, they still indicate  a curious
boundary effect in local inflationary scenarios.

\end{abstract}

\maketitle

\section{Introduction}

A remarkable property of the observed universe is its large scale homogeneity and isotropy. For instance, the temperature of the cosmic microwave background (CMB) is isotropic  one part in $10^{5}$ across the sky. This smoothness appears to contradict with a generic  big-bang since in that case one would expect to see imprints of different exotic objects on CMB. In particular, white holes should have been created in big-bang because a common big-crunch necessarily contains black-holes and big-bang can be viewed as a time reversed big-crunch. An important open problem in  modern cosmology is to explain why the universe appeared in this very special state.

It is generally claimed that the standard cosmological model cannot explain  the isotropy of  CMB, since in this model most of the  cosmic photons we observe today originates from  causally disconnected regions in space and thus thermalization cannot take place to yield a uniform temperature.  A plausible way to solve this difficulty (and others like flatness and monopole  problems) is to modify the standard model by assuming an early period of accelerating expansion, called inflation. Although the details of how the expansion is achieved depend on the model one considers, in general there is a scalar field whose potential energy acts like an effective  cosmological constant yielding a de  Sitter phase. By the huge exponential expansion encountered in this period all inhomogeneities are smoothed out and the spatial sections are  flattened out. The temperature drops enormously at the end of inflation. However, following a reheating process  the temperature again raises and  one ends up with a radiation dominated universe.

Since in a cosmological model without inflation causality precludes thermalization, it seems impossible to justify  special properties  of CMB. However, as  pointed out by Penrose  (see e.g. \cite{penrose}), it is paradoxical to view the present  specialness as a feature that should be explained and use thermalization as  the main mechanism to account for it, since this already means that the universe should have been more special in the past. Namely, one should try to understand possible cosmic origins of the second law of thermodynamics instead of using it in explaining the current state of the universe. Furthermore, without understanding Planck scale physics it is impossible to determine the degree of anisotropy in the absence of  inflation. Therefore it is not possible to view the observed isotropy as a feature contradicting with the standard model.

Although inflation is proposed as a model explaining the homogeneity and isotropy of the observed universe, one usually assumes a homogeneous and isotropic distribution of a scalar field (having a suitable potential) to realize it. To resolve this  conflict one supposes that inflation occurred in a homogeneous and isotropic local patch in a larger (radiation dominated) Freedman-Robertson-Walker-Lemaitre (FRWL) spacetime. One then wonders whether inflation is 'natural' in this set up. In some cases, it is possible to prove a cosmic no hair theorem which asserts that  a positive cosmological constant eventually dominates the cosmic evolution irrespective of initial conditions and matter content \cite{wald1,str} (see also \cite{rend}).  However, this theorem does not say too much about feasibility of local inflation which is not derived by  a pure cosmological constant. Although chaotic inflation appears to be an attractive point in simple scalar field models with  suitable potentials \cite{linde}, the issue of initial conditions for inflation is not yet settled (for a review see e.g. \cite{review}). As illustrated  in \cite{wald2},  without having a natural measure in the space of initial conditions, it is impossible to assign  a probability for inflation. Remarkably, a measure which yields an exponentially suppressed probability has been proposed recently in \cite{gibbons}. Moreover, naive arguments indicate that the entropy of the pre-inflationary patch is much lower than a  patch with a typical big-bang,  which shows that among randomly chosen initial conditions the probability of seeing inflation should be negligible (see e.g. \cite{carroll}). On the other hand, there are some obstructions in embedding a "small" inflating region into a FRWL universe imposed by the propagation of null geodesics \cite{hom1}. A conceivable way to remedy this obstruction is proposed in  \cite{hom2}.

Even though the original motivations mainly arise from the "shortcomings" of the standard model, one of the main successes of the inflationary paradigm turns out to be its ability  to produce  a scale free spectrum of density perturbations appropriate for the formation of structure in the universe. According to the standard picture, density perturbations originate from quantum fluctuations,  which are spontaneously created in  de Sitter space and amplified  over superhorizon scales. In this paper, we point out that in  local inflationary  models there also exists (quantum) perturbations which inevitably enter into the inflating region from outside, whose power spectrum is not scale invariant.  These perturbations can spread out a horizon distance from the boundary and they may influence structure formation in that region.

The plan of the paper is as follows. In the next section, we state our main argument. In section \ref{sec3}, we argue that due to the causal structure of local embedding some outside perturbations inevitably enter  into the inflating region and we illustrate how our main argument is realized in the thin-shell approximation. In section  \ref{ana}, we make a further check of scale dependence by matching the interior and the exterior modes analytically. We briefly conclude in section \ref{sec4}.

\section{The basic argument}\label{sec2}

Let us start by summarizing the well-known mechanism for the generation of scale free spectrum of density perturbations in inflation. Consider  a scalar field $\phi$ obeying the Klein-Gordon equation  $\nabla^2 \phi=0$ in a spatially flat Robertson-Walker spacetime
\be
ds^2=-dt^2+a(t)^2\left(dx^2+dy^2+dz^2\right).
\ee
For a plane wave mode characterized by a comoving  wavenumber $k$ there are two important propagation limits. When the proper wavelength  $a/k$ is much smaller than the Hubble radius $R_H=1/H$, the mode will behave like an ordinary harmonic oscillator with negligible damping. In that case, the ground state of the oscillator is a Gaussian wave function with spread given by (see appendix \ref{ap0})
\be
\D\phi^2_k=\fr{1}{2a^2 k}. \label{amp}
\ee
On the other hand, when the wavelength $a/k$ is much larger than $R_H$, the oscillator is overdamped and the fluctuation amplitude $\D\phi^2_k$ freezes out.

During inflation  $R_H$ is constant but the physical wavelengths increase exponentially. Therefore, a mode can evolve from an underdamped oscillator  to an overdamped oscillator, but the opposite is not possible. One assumes that  throughout  inflation quantum modes were born in their ground states with wavelengths much smaller than the Hubble radius and evolve adiabatically with the spread given by \eq{amp} until their  wavelengths become equal to $R_H$. For a mode with wavenumber $k$ this happens when $a/k=R_H$ and $\D\phi^2_k=H^2/(2k^3)$. Later on, the oscillator becomes overdamped, the wavelength leaves the horizon and the amplitude freezes out. The corresponding power spectrum $P(k)\equiv k^3 \D\phi^2_k$  is independent of $k$, i.e. $P(k)$ is  scale free.  After inflation ends, the Hubble radius  grows more rapidly than the scale factor $a$, and thus the wavelengths created during inflation reenter the horizon eventually.

Let us point out that a superhorizon perturbation is still {\it propagating}  even though  its amplitude freezes out. Consider, for instance, a mode in de Sitter space whose metric in conformal time $t_c<0$ becomes
\be
ds^2=(Ht_c)^{-2}\left(-d t_c^2+dx^2+dy^2+dz^2\right).
\ee
It is well known that the exact solution of $\nabla^2 \phi=0$ for a plane wave perturbation can be written as (see e.g. chapter 7 of \cite{linde2})
\be\label{mod}
\phi_k=\fr{H}{\sqrt{2k}}\left[-t_c+\fr{i}{k}\right]\,e^{i(-kt_c\pm\vec{k}.\vec{x})},
\ee
where the overall normalization is fixed by the harmonic oscillator commutation relation
\be
\dot{\phi}_k^*\phi_k-\dot{\phi}_k\phi_k^*=i (Ht_c)^2.
\ee
Note that $\phi_k=\s_k(-Ht_c)$ where $\s_k$ is the canonical harmonic oscillator excitation. The subhorizon and superhorizon limits correspond to $k|t_c|\gg1$ and $k|t_c|\ll1$, respectively. Although the behavior of the amplitude differs in the two limits, the phase dependence implies  propagation along the null characteristic hypersurfaces. This should somehow be anticipated since these are nontrivial  solutions of $\nabla^2 \phi=0$. Therefore, a mode being frozen out does not mean it is non-propagating, this simply signifies that its amplitude becomes time independent.

In visualizing local inflation, one usually imagines that the inflating region is exponentially expanding into the ambient spacetime.  Although it might be possible to picture inflation using physical coordinates, this may create some confusion in subtle considerations. It is a lot  safer to consider comoving coordinates in describing cosmological phenomena. In these coordinates the region itself does not necessarily expand. Inflation is realized by an exponentially growing metric in a (possibly fixed) comoving region.

Another crucial point is that in general the inflating patch cannot be considered as a closed system. Although in some cases there may form "isolated" inflating universes  \cite{ts3}, we will see in the next section that  even in this situation the inflating region is not  causally disconnected.  Therefore,  some exterior information can penetrate inside. Indeed, one would expect a tendency of flowing from the  surrounding region having positive pressure into the  inflating region with negative pressure (see e.g. \cite{mwu}). It is even possible for this flow to stop inflation, however we will assume that this is not the case.

\begin{figure}
\centerline{
\includegraphics[width=6.0cm]{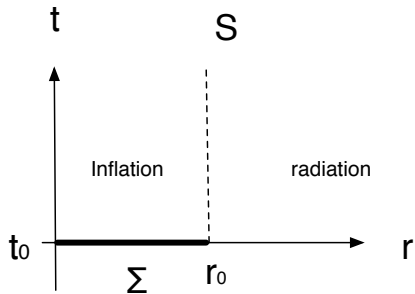}}
\caption{A spacetime sketch of local inflation in {\it comoving coordinates}. The initial  conditions at time $t_0$ is suitable for a spherical region $\S$ of radius $r_0$ to inflate in a radiation dominated FRWL space. Both regions are separated by the hypersurface $S$.}
\label{fig1}
\end{figure}

Consider, for instance, a radiation dominated FRWL universe. Let us assume that at time $t_0$ a local spherical region $\S$ of radius $r_0$ starts inflating for some reason (see figure \ref{fig1}), where $r$ and $t$ are  comoving coordinates.  Let $S$ be the hypersurface that separates the inflating region from radiation dominated space. In general $S$ can be a thick hypesurface, i.e. a four dimensional submanifold, but we assume that a  thin wall approximation is valid.  Denoting the scale factors as $a_I$ and $a_R$, respectively, one can normalize coordinates such that  $a_I(t_0)=a_R(t_0)=1$ and thus
\be\label{sf}
a_I=e^{H(t-t_0)},\hs{4}a_R=\left(\fr{t}{t_0}\right)^{1/2},
\ee
where $H$ is the Hubble parameter in the inflating region. The corresponding Hubble radii are $R_I=1/H$ and $R_R=2t$. Causality requires that  $r_0$ should be  smaller than the Hubble radius of the FRWL space at time $t_0$, i.e. $2t_0> r_0$. Moreover, it is known that for stability vacuum energy should dominate over a region larger than the inflationary horizon, $r_0 > 1/H$  (see e.g. \cite{hom1}).
In any case there is no reason to expect a large hierarchy between these three scales, i.e.
\be\label{hie}
2t_0\sim r_0\sim \fr{1}{H}.
\ee

Let us think about a (quantum) scalar perturbation that appeared at time $t$ in the FRWL space whose physical wavelength $\l_p$ is smaller than the Hubble radius
\be\label{lp}
2t>\l_p.
\ee
This can be represented as an underdamped harmonic oscillator that was born in its ground state whose wavefunction has the spread
\be
\D\phi^2_k=\fr{1}{2a_R^2 k} \label{ampR},
\ee
where $k$ is the corresponding comoving wavenumber $k=a_R/\l_p$. The mode can be assumed to evolve adiabatically with \eq{ampR} in the geometric optics limit, i.e. on the null rays of FRWL space. Moreover it cannot turn into an overdamped oscillator  since the Hubble radius is growing faster than the physical wavelength.

\begin{figure}
\centerline{
\includegraphics[width=6.5cm]{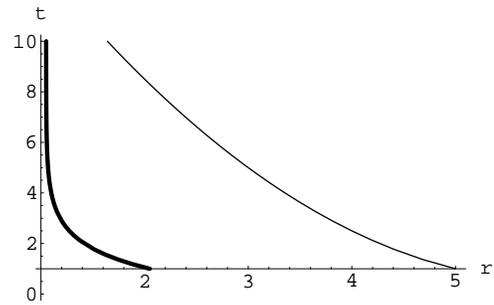}}
\caption{Left moving radial null rays of de Sitter (thick line) and radiation dominated FRWL space (thin line) drawn in the same comoving coordinates in units where $H=1$. Initially the Hubble constants are chosen to be equal to each other.}
\label{fig2}
\end{figure}

Assume now that  (a portion of this) mode (containing a single pulse of one wavelength) enters into the inflating space at time $t^*$. Since we are using same comoving coordinates in both regions, the comoving wavenumber $k$ does not change. Therefore, the physical wavelength of the mode as it enters into the inflating region is amplified by
\be\label{lpi}
\l^*=\fr{a_I(t^*)}{a_R(t^*)}\, \l_p.
\ee
From \eq{hie} and \eq{lp}, we see that there is no hierarchy between $\l_p$ and $1/H$. Therefore,  unless $\l_p$ is exponentially small compared to $1/H$,  \eq{lpi} implies that
\be
\l^*\gg \fr{1}{H}.
\ee
Consequently, this mode should freeze out in the inflating region. From \eq{ampR}, the fluctuation spectrum can be found as
\be
\D\phi^2_k=\fr{1}{2a^2_R(t^*) k} \label{ampI},
\ee
which acquire  a {\it scale dependent}  power spectrum with $P(k)\sim k^2$.

Although in the above argument we assume that the mode enters into the inflating region at a single time $t^*$ and amplified immediately, this actually happens in a time interval proportional to the wavelength of the mode (see next section). Therefore, it is important to note that the amplification \eq{lpi} does not occur immediately.  Moreover, as it  will be seen  in the next section, it is not always possible to use same comoving coordinates in both regions. As a result, the comoving wavenumber may alter during crossing.  As we will show, however, these complications do not change the main conclusion that outside perturbations acquire a scale dependent power spectrum.

As these perturbations propagate into the inflating region, they can spread out a comoving radial distance $1/H$ from the boundary, which is the inflationary horizon distance (this is due to the behavior of the radial null lines in de Sitter space, see e.g. figure \ref{fig2}). Therefore, they cannot reach out an observer whose comoving distance is larger than $1/H$ from the boundary. In models where $r_0\gg 1/H$, the probability of seeing a typical observer (us) near the boundary is small and thus these modes  would most likely be invisible to us to  produce an observational effect.

\section{A toy model and the crossing of a scalar perturbation along the boundary}\label{sec3}

As pointed out above, since the inflating region has negative pressure as oppose to the exterior with positive pressure, there should be a tendency of flowing into the inflating patch. This tendency can easily be seen by comparing radial null lines of each space pictured in the same comoving coordinates; it turns out that  (when the Hubble constants are comparable) the null curves in de Sitter space is much more vertically aligned compared to the  ones in a radiation dominated spacetime (see figure \ref{fig2}). Therefore, information spreads a lot faster in the radiation region.

Whether outside perturbations can enter into the inflating space or not depends on the causal structure of the hypersurface $S$ separating two regions. To determine $S$ one should study the coupled dynamics of  the metric and the scalar field driving inflation. This problem can be addressed in the thin-shell approximation \cite{thin} and indeed the evolution of an  inflating region (the false vacuum bubble) in a  cosmological background has been studied well in this context (see e.g. \cite{ts1,ts2,ts3,ts4}).  In the thin-shell approximation, two spaces are glued over the boundary $S$ describing the history of the bubble, which is a {\it timelike} hypersurface in both regions. The induced metric is taken to be continuous over $S$ and the difference of the extrinsic curvatures is equal to the energy momentum tensor on the shell.

\begin{figure}
\centerline{
\includegraphics[width=5.0cm]{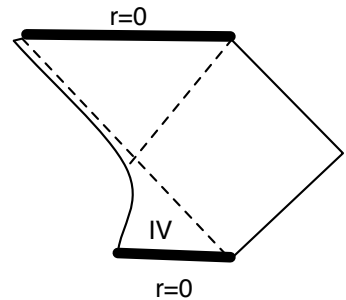}}
\caption{The sketch of a solution for $S$ given in  \cite{ts3} as seen by an outside observer, which yields an "isolated" inflating universe on the left.}
\label{fig6}
\end{figure}

Consider, for instance, a well known example from \cite{ts3},  where the evolution equations describing  a   spherical false vacuum bubble in a true vacuum region are explicitly solved. It is argued in \cite{ts3} that at least for a range of initial conditions there forms "isolated" closed inflating universes in this setup.  A solution, as seen by an outside observer, is pictured in figure  \ref{fig6}. The outside region should uniquely be described by the black hole metric as given by the Birkhoff's theorem. Here, we see that the boundary is formed by a timelike curve starting from the past singularity at $r=0$ and  extending throughout  future null boundary. As pointed out in \cite{ts3}, if the evolution is foliated on suitable constant time hypersurfaces, one sees a closed inflating universe that is detaching.\footnote{A critical objection to this picture is raised in \cite{obj}  by indicating  that  the solution involves the past of a bifurcating Killing horizon which is known to be unstable. The stability issue in the thin shell formulation of false vacuum bubbles is also considered in \cite{stab}.} However, it is clear that the inflating region is not causally disconnected from the black hole region. Indeed all left moving perturbations created in region IV in the figure \ref{fig6}  enter into the de Sitter space and according to the argument presented in the previous section they acquire a scale dependent power spectrum.

Since in the thin-shell formulation $S$ is always taken to be a timelike hypersurface, it is inevitable that some exterior perturbations (propagating along null geodesics) enter into the inflating space. To explain how the main mechanism discussed in the previous section works in this context, we construct  an explicit toy example in the appendix \ref{ap1}, where an inflating region and a flat Minkowski space are  glued over the worldline of a spherical bubble. The line element in each patch is taken as
\be
ds^2 = - dt^2+ a(t)^2\left(dr^2+r^2 d\O_2^2\right),
\ee
where $a=1$ in the flat space and $a=e^{Ht}$ in the inflating region. For a spherically symmetric configuration, the two spheres in each region can be identified by the bubble itself. However, $r$ and $t$ coordinates are in general discontinuous.   From appendix \ref{ap1} (see \eq{asol1} and \eq{asol2}) we quote the equations describing the boundary: in the flat space $S$ is given by
\be\label{sol1}
r=\fr{1}{\a}\cosh(\a \t),\hs{4} t=\fr{1}{\a}\sinh(\a \t),
\ee
and in the inflating region it reads
\bea
&&r=\fr{r_0\cosh(\a\t)}{|\sinh(\a\t)+\fr{4H^2-\s^2}{4H\s}|},\label{sol2}\\
&&Ht=\ln(|\sinh(\a\t)+\fr{4H^2-\s^2}{4H\s}|)-\ln(\a r_0).  \nn
\eea
Here $r_0$ is an integration constant, $\s>0$ is the constant energy density on the boundary,  $\a$ is given by
\be
\a=\fr{H^2}{\s}+\fr{\s}{4},
\ee
and $\t$ is the proper time. To avoid any complication one can assume that $H$, $\s$ and thus $\a$ have the same order of magnitude.

As pointed out in the appendix \ref{ap1}, to solve the junction conditions in this simplified set up one should  identify flat space as the interior and de Sitter space as the exterior patches. Therefore, these equations actually describe the evolution of a true vacuum bubble in an inflating space.

\begin{figure}
\centerline{
\includegraphics[width=5.5cm]{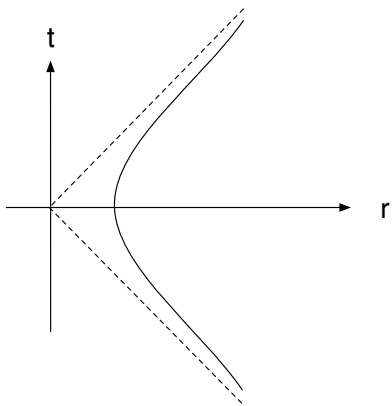}}
\caption{The worldline of the bubble boundary as seen by an observer in the flat space.}
\label{fig9}
\end{figure}

\begin{figure}
\centerline{
\includegraphics[width=5.5cm]{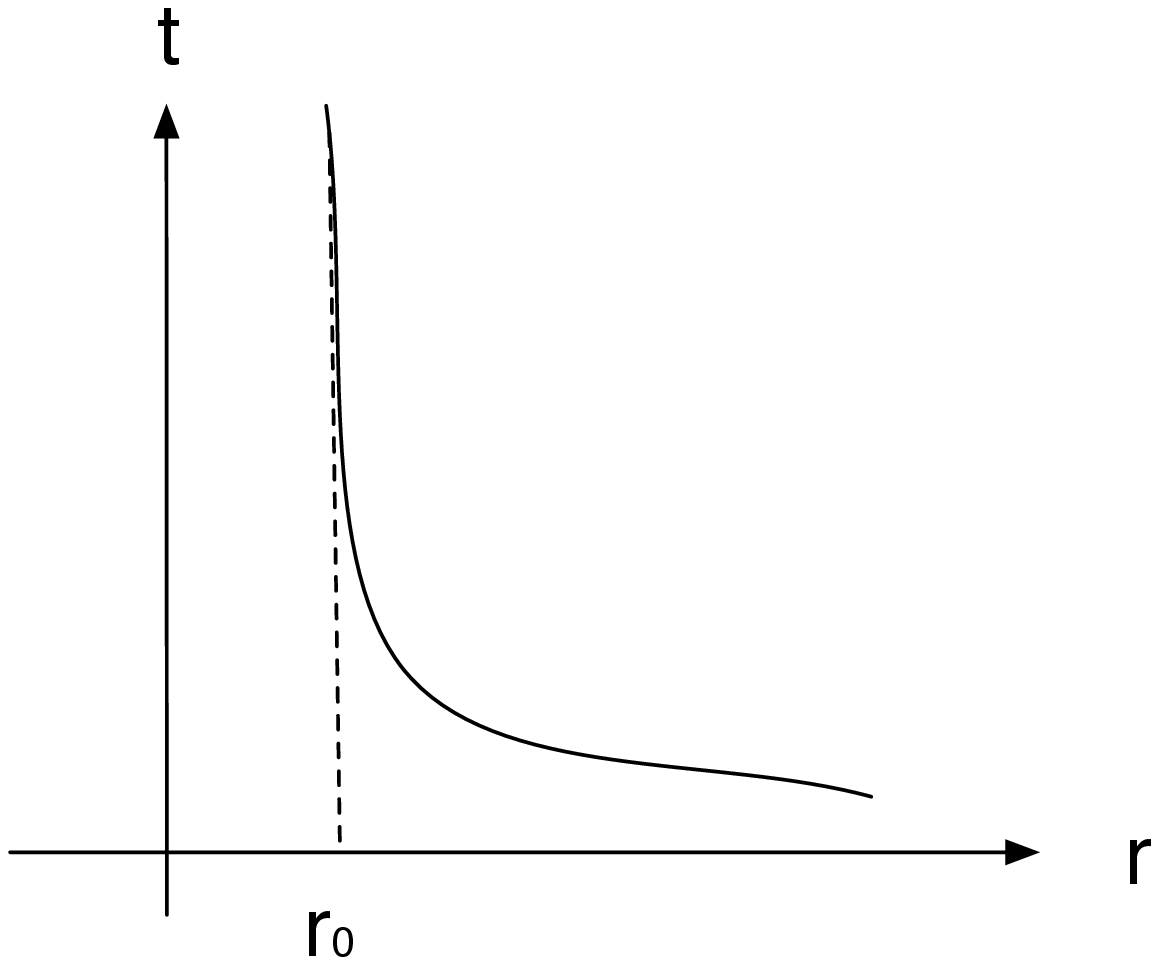}}
\caption{The worldline of the bubble boundary  as seen by an observer in de Sitter space when $4H^2-\s^2\leq0$.}
\label{fig7}
\end{figure}

\begin{figure}
\centerline{
\includegraphics[width=5.5cm]{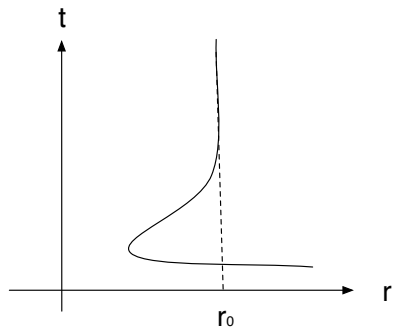}}
\caption{The worldline of the bubble boundary  as seen by an observer in de Sitter space when $4H^2-\s^2>0$.}
\label{fig8}
\end{figure}

The flat space  trajectory  \eq{sol1} corresponds to the worldline of a particle moving with constant acceleration $\a$ equivalent to a hyperbola (see figure \ref{fig9}). To understand the bubble solution in de Sitter space  better, we note that there is a singularity at  $\t=\t_s$, where the denominator of \eq{sol2} vanishes. One should restrict $\t>\t_s$ since in that range $t'>0$, where prime denotes differentiation with respect to $\t$. Asymptotically as $\a\t\to\infty$, $r\to r_0$. The sign of $r'$ is always negative when  $4H^2-\s^2\leq0$. If   $4H^2-\s^2>0$, there is a turning point after which $r'>0$.  The corresponding paths are plotted in figures \ref{fig7} and \ref{fig8}.

\begin{figure}
\centerline{
\includegraphics[width=5.5cm]{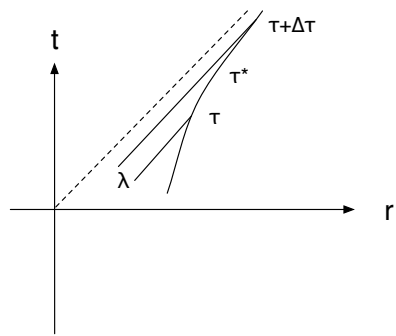}}
\caption{A right moving spherical perturbation of wavelength $\l$ crossing the hypersurface $S$.}
\label{fig10}
\end{figure}

Let us now determine the power spectrum of perturbations which are spontaneously created in the true vacuum region and passing over the inflating space. Consider a properly normalized right moving spherical perturbation of wavelength $\l$ in flat space
\be\label{modfl}
\phi_k=\fr{1}{\sqrt{2k}}\fr{e^{ik(r-t)}}{r},
\ee
where $k=2\pi/\l$. When a  single pulse of {\it one} wavelength extending in between the null interval $r-t$ and $r-t+\l$ reaches the boundary (see figure \ref{fig10}), it overlaps with a portion in the proper time interval ($\t,\t+\D\t)$. We assume that $\D\t\ll\t$, i.e. the crossing time is small compared to the whole history. Therefore, the  approximate time of passing can be defined as $\t^*=\t+\D\t/2$. To make sure that the inflating region already expanded enough, we focus on the perturbations entering at late times, i.e. $e^{\a\t}\gg1$. Using \eq{sol1}, one finds that
\be\label{l}
\l=\fr{2}{\a}e^{-\a\t^*}\sinh(\a\D\t/2).
\ee
Given any $\t^*$ this equation relates  $\D\t$ to $\l$.

\begin{figure}
\centerline{
\includegraphics[width=5.5cm]{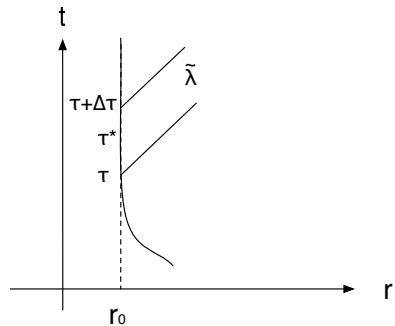}}
\caption{The perturbation in figure \ref{fig10} exits  as a spherical wave of comoving wavelength $\tilde{\l}$ in de Sitter space.}
\label{fig11}
\end{figure}

We would like to match this pulse with a spherically symmetric perturbation in de Sitter space. The mode
which has a comoving wavenumber $\tilde{k}$ can be written as
\be\label{modds}
\phi_{\tilde{k}}=A\,\fr{H}{\sqrt{2\tilde{k}}}\fr{e^{i\tilde{k}(r-t_c)}}{r} \left[-t_c+\fr{i}{\tilde{k}}\right],
\ee
where $t_c$ is the conformal time defined by
\be
Ht_c=-e^{-Ht}.
\ee
Since this is not an oscillator mode  spontaneously created in de Sitter space, there is a  freedom in normalization which is indicated by the  unknown  constant $A$ in \eq{modds}.  Using \eq{sol2},  the comoving wavelength $\tilde{\l}$ of a pulse extending in between $r-t_c$ and $r-t_c+\tilde{\l}$ can be related to $\t^*$ and $\D\t$ as (see figure \ref{fig11})
\be\label{tl}
\tilde{\l}=\fr{4\a r_0}{H}e^{-\a\t^*}\sinh(\a\D\t/2).
\ee
Comparing now \eq{l} and \eq{tl}, we obtain
\be\label{ktk}
k=\fr{2\a^2 r_0}{H}\,\tilde{k}.
\ee
From \eq{ktk}, it is easy to see that if the wavelength in the flat space obeys
\be\label{cond}
\l>\fr{2\pi}{\a}e^{-\a\t^*},
\ee
the corresponding mode in de Sitter space satisfy $\tilde{k}|t_c(\t^*)|<1$. Therefore, these excitations become superhorizon perturbations in the inflating region. Matching the interior and exterior scalar modes up to phase on the boundary one further finds
\be\label{ampA}
A=\sqrt{\fr{2r_0}{H}}e^{-\a\t^*}\tilde{k}.
\ee
Due to this $\tilde{k}$ dependence, \eq{modds} is not canonically normalized and  the overall amplitude gives  $\D\phi^2_{\tilde{k}}\sim 1/\tilde{k}$. The corresponding power spectrum becomes scale dependent $P(\tilde{k})\sim \tilde{k}^2$, which is consistent with the results obtained in the previous section.

At this point one may worry that the amplitude \eq{ampA} is exponentially suppressed. However, this suppression arises from $1/r$ dependence of  \eq{modfl}, i.e. this behavior is specific for a spherical wave created at the origin and does not indicate a generic feature.  It is also interesting to note that \eq{l} and \eq{cond} implies
\be
\sinh(\a\D\t/2)>\pi.
\ee
This shows there is a minimum time width of boundary crossing for perturbations which become superhorizon in the inflating region.

In general, one has to include a scattered left moving wave in flat space corresponding to reflection from the boundary (see next section). The reflection can be ignored if the radius of curvature of the boundary (in our case $1/\alpha$) is much larger than the wavelength $\l$. From \eq{cond} we see that for $e^{\a\tau^*}\gg1$ there is a band of wavelengths obeying both \eq{cond} and $\l\ll 1/\alpha$. Thus for these modes the reflection can be neglected  and the above conclusions must hold.

\section{Analytical matching}\label{ana}

In the previous section, in matching the interior and the exterior modes over the boundary, we have simply considered a pulse of one wavelength in both regions. Although this has been sufficient  to  illustrate our argument, to be more precise, one should actually glue the whole solutions.  To address this problem, we first note that the  Klein-Gordon equation can be expressed in terms of the derivative operator $D_\m$ of the induced metric  on $S$ and the normal vector $n^\m$ as
\be
\nabla^2\phi=D^2\phi+Kn^\m\del_\m\phi+n^\m\del_\m(n^\n\del_\n\phi)-(n^\m\nabla_\m n^\n)D_\n\phi.
\ee
Since $K$ is discontinues but does not involve a delta function,  this equation implies that two fields $\phi^-$ and $\phi^+$, which are obeying $\nabla^2\phi^\pm=0$ in respective regions, can be glued
on the boundary if they satisfy
\bea
\phi^-&=&\phi^+,\label{cont1}\\
n^\m\del_\m\phi^-&=&n^\m\del_\m\phi^+ \label{cont2}.
\eea
We now use these junction conditions to study the crossing of a perturbation.

In our case, there is an incoming wave which is the properly normalized, right moving, spherical perturbation
\be\label{modin}
\phi_{in}=\fr{1}{\sqrt{2k}}\fr{e^{ik(r-t)}}{r}.
\ee
The reflected and the transmitted modes   can be taken as the general, spherically symmetric, left and right moving waves in flat and de Sitter spaces, respectively. These can be written as
\be\label{mods}
\phi_{ref}=\fr{g(1/(r+t))}{r}, \hs{8} \phi_{tr}=\fr{f'(r-t_c)t_c}{r}+\fr{f(r-t_c)}{r},
\ee
where $f,g$ are arbitrary functions and prime denotes differentiation with respect to the argument.
The general solution in de Sitter space  can be obtained from \eq{modds} by superposition. Two conditions  \eq{cont1} and \eq{cont2} can be used to determine two unknown functions $f$ and $g$. Note that $\phi^-=\phi_{in}+\phi_{ref}$ and $\phi^+=\phi_{tr}$.

We consider the late time matching of the solutions, i.e. we assume $e^{\a\t}\gg1$. From \eq{sol1} and \eq{sol2}, the components of the normal vector $n^\m$ can be found in flat space as
\be
n^t=n^r=\fr{e^{\a\t}}{2},
\ee
while in de Sitter space they read
\be
n^t=\fr{4H^2-\s^2}{4H\s},\hs{8}n^r=\fr{2r_0\a^2}{H}e^{-\a\t},
\ee
where we only keep the leading order contribution in an expansion in $e^{-\a\t}$. We define  a (dimensionfull)  variable $x\equiv \a e^{-\a\t}$. In units where $\a=1$ we have $x\ll1$. To first order in $x$, the continuity of the modes \eq{cont1} implies
\be\label{c1}
\sqrt{\fr{2}{k}}\,x\,e^{ikx/\a^2}+2\,x\,g=\fr{1}{r_0}f-\fr{2}{H}xf',
\ee
where prime denotes differentiation with respect to the argument and
\be\label{arg}
g=g(x),\hs{10}f=f(r_0+\fr{2r_0}{H}x).
\ee
On the other hand, the continuity of normal derivatives \eq{cont2} gives
\bea
\left(i\fr{\sqrt{2k}}{\a}x-\fr{4\a}{\sqrt{2k}}\right)e^{ikx/\a^2}-2\a x g'-2\a g\nn\\
=-\fr{2r_0\s }{H^2}xf''+\fr{2\a}{H}f'-\fr{2\a}{r_0H}f.\label{c2}
\eea
Solving $g$ from \eq{c1} and using it in \eq{c2} we obtain the following second order equation for $f$:
\bea
\fr{2r_0(\s+2\a)}{H^2}xf''&-&\fr{2\a}{H}f'+\fr{2\a}{r_0H}f\nn\\
&=&\left(\fr{2\a}{\sqrt{2k}}-2i\fr{\sqrt{2k}}{\a}x\right)e^{ikx/\a^2}. \label{f}
\eea
The right hand side of this equation is the source term related to the incoming wave.

It is possible to solve \eq{f} exactly. The homogenous solution involves two Bessel functions and the particular solution can be obtained by variation of parameters. We need to keep the unique particular solution supported by the source incoming wave.  However,
\eq{f} has corrections in powers of $x$, i.e. both the functions multiplying $f$ and the source in the right hand side are just the leading order contributions. Therefore, only the first term in the expansion of the exact solution in $x$ can be trusted.

Since the source term in \eq{f} is oscillating, the solution can be written as $f=e^{ikx/\a^2} F(x)$, i.e. $f$ will oscillate exactly the same way as the source.  From \eq{arg} we then find
\be
f(y)=e^{ikHy/(2\a^2 r_0)}...\, ,
\ee
thus the comoving wavenumber $\tilde{k}$ in de Sitter space can be identified as $\tilde{k}=kH/(2\a^2r_0)$, which is precisely the relation \eq{ktk} derived in the previous section.

Let us now determine the solution of \eq{f} in the short and long wavelenght limits. To avoid any complication we assume that  $H$, $\s$ (and thus $\a$) and $1/r_0$ have the same order of magnitude.

We start with the short wavelength limit $kx/\a^2\gg1$, where the first term in the right hand side of \eq{f} can be neglected and the solution can be expanded as
\be
f=A\,e^{ikx/\a^2}(1+{\cal O}(x)).
\ee
In \eq{f}, $f''$ term becomes much larger than the two other terms in the left hand side and the constant $A$ can be fixed as
\be\label{ak}
A=i\fr{8\a^3r_0}{(\s+2\a)}\fr{1}{k\sqrt{2k}}.
\ee
Here $k$ dependence is crucial. Using $f$ in \eq{mods}, we find
\be
\phi_{tr}=\left[\fr{4\a e^{-iHk/2\a^2}}{(\s+2\a)}\right]\fr{H}{\sqrt{2k}}\fr{e^{i\tilde{k}(r-t_c)}}{r}\left(-t_c+\fr{i}{\tilde{k}}\right).
\ee
The constant in the square brackets is an order unity constant. Comparing with \eq{mod} and nothing that $\tilde{k}$ and $k$ are nearly equal to each other (recall that we assume $H$, $\a$ and $1/r_0$ have the same order of magnitude), we see that the transmitted wave is a properly normalized mode in de Sitter space. Indeed, in the limit that we consider $1/\tilde{k}$ term can be neglected compared to $t_c$ in the parentheses, which shows that this is a subhorizon mode. Therefore, perturbations which have short wavelengths in flat space can be matched by subhorizon modes in the inflating region which acquire a scale free power spectrum. This is consistent with   the results of the section \ref{sec2} since the amplification \eq{lpi} is not enough to push the physical wavelengths of these modes out of the inflationary horizon.

Consider now the long wavelength limit $kx/\a^2\ll1$. In this limit the second term in the right hand side of \eq{f} can be ignored, the exponential can be set to unity and the solution can be expanded as
\be
f=A(1+{\cal O}(x)).
\ee
The amplitude can be fixed from \eq{f} as $A=r_0H/\sqrt{2k}$ which yields  a spherical, superhorizon mode in de Sitter space
\be
\phi_{tr}= \fr{r_0H}{\sqrt{2k}r}.
\ee
However the normalization is different than the normalization of the canonical superhorizon perturbation, which is $1/k\sqrt{2k}$ . This gives a scale dependent power spectrum with $P(k)\sim k^2$, which is consistent with the results presented in the previous sections.

\section{Conclusions}\label{sec4}

Inflation offers a successful way of generating a scale free spectrum of density perturbations consistent with cosmological observations. According to the general view, all classical inhomogeneities  are smoothed out by the exponential expansion, leaving only room for quantum fluctuations. Naturally, quantum modes are assumed to be born as harmonic oscillators  in their ground states, which have physical wavelengths smaller than the inflationary horizon. The wavelengths are then pushed to superhorizon sizes by expansion and their amplitudes freeze out. After inflation ends, they reenter the horizon in a radiation dominated phase as classical perturbations  seeding cosmic structure.

The success of the above mechanism profoundly  depends on the very special properties of  the generation and the propagation of quantum fluctuations in de Sitter space.  Indeed,  even in de Sitter space the well known ambiguity in the choice of an invariant vacuum may alter some of the predictions of the theory (see e.g. \cite{dan}). In this paper, we point out that in local inflationary models the causal embedding of an inflating patch into an ambient spacetime generally allows outside perturbations to enter into the inflating region. It is clear that these perturbations may not have  a scale-free power spectrum.

In this paper  we show that quantum fluctuations entering into a local inflating patch have a scale dependent power spectrum. In section \ref{sec2}, we first give a general argument and in sections \ref{sec3} and \ref{ana}  we illustrate it by an explicit toy example in the thin-shell approximation. Although the set up studied in this context is artificial and cannot be considered as a part of a  realistic scenario, it nicely touches to the salient points raised in section \ref{sec2}. It would be interesting to construct a numerical or an asymptotic false vacuum bubble solution in a radiation dominated FRWL space and verify that scale freeness is spoiled for the quantum modes crossing into the inflating region, as suggested by the basic argument of section \ref{sec2}.

One may think that  the above conclusions are avoided in models where the inflating patch detaches from the ambient spacetime. We consider one such example from \cite{ts3} and indicate that the patch is still causally connected to the surrounding region. Indeed, in the thin-shell approximation causal detachment is impossible since the boundary in between the two regions is taken to be a timelike hypersurface in both sides.

The key issue is then to determine whether causal disconnection is possible in local inflationary models at all. One of the nicest features of inflationary paradigm is that it uses classical general relativity in determining the evolution of the metric and thus its predictions are not sensitive to the quantum gravitational corrections.  In general relativity, on the other hand,  causal detachment implies the existence of a real event horizon.  To our knowledge black holes are only examples where observer independent causal separations take place. In some models using chaotic inflation, it is argued that after big bang (or after universe emerges from a spacetime foam) regions suitable for inflation expands and others recollapse.  In such local models it seems that the above conclusions are avoided. However, it is not clear to us how this picture can be verified without understanding the end point of a collapse, which requires a detailed knowledge of quantum gravity. Indeed, even in semiclassical gravity black holes radiate energy and thus it appears that  a complete causal detachment is not possible.

Finally, these findings do not imply a disagreement with observations in local inflationary models. To claim a discrepancy, one should first make sure that the contribution of outside quantum perturbations to structure formation is not negligible. Moreover, as pointed out previously, depending on the size of the initial inflating patch and the location of the observer inside, the outside perturbations may not be able to reach the observer since in de Sitter space there are particle horizons and an observer is shielded by a horizon. In any case, our results indicate that the boundary effects should be considered as an issue in local inflation.

\appendix

\section{Scalar field quantization in a FRWL space} \label{ap0}

In this appendix we review the quantization of a real, massless scalar field in a FRWL spacetime with the metric
\be\label{metilk}
ds^2=-dt^2+a(t)^2\left(dx^2+dy^2+dz^2\right).
\ee
The canonical action can be taken as
\be\label{act}
S=-\fr{1}{2}\int\sqrt{-g}\,g^{\m\n}\,\del_\m\phi\,\del_\n\phi,
\ee
which gives $\nabla^2\phi=0$.  Since in a free field theory different Fourier modes decouple, we consider a single excitation with a fixed  comoving wave-vector $\vec{k}$:
\be
\phi=\phi_k(t) e^{i\vec{k}.\vec{x}}+\phi_k^\dagger(t) e^{-i\vec{k}.\vec{x}},
\ee
where the field equations imply
\be\label{fk}
\ddot{\phi}_k+3H\dot{\phi}_k+\fr{k^2}{a^2}\phi_k=0.
\ee
From \eq{metilk} and \eq{act} the momentum conjugate to $\phi$ becomes $P_{\phi}=a^3\dot{\phi}$. The quantization can be achieved by imposing
\be\label{commt}
[\phi,P_{\phi}]=i,\hs{5}\phi_k |0>=0,
\ee
where $|0>$ is the vacuum as seen by a comoving observer.

For a subhorizon mode $k/a\gg H$ the friction term in \eq{fk} can be ignored and an approximate solution for $\phi_k$ can be found as
\be
\phi_k=a_k\,e^{-ikt/a}.
\ee
The commutator in \eq{commt} then implies
\be
[a_k,a^\dagger_k]=\fr{1}{2ka^2}.
\ee
Therefore, $a_k$ and $a_k^\dagger$ are creation and annihilation operators and the system is equivalent to the harmonic oscillator. One can easily calculate  the two-point function or the spread of the Gaussian ground state wave-function as
\be
\D \phi_k^2\equiv<0|\phi^2|0>=\fr{1}{2a^2 k}.
\ee

\section{Derivation of the bubble solution}\label{ap1}

The aim of this appendix is to obtain an explicit equation for the hypersurface $S$ which is separating a flat Minkowski space from an inflating region. We start by considering the actual problem, i.e. a false vacuum bubble in a FRWL space. The metrics in both  regions can be written as
\be
ds_\pm^2 = - dt_\pm^2+ a_\pm(t_\pm)^2\left(dr_\pm^2+r_\pm^2 d\O_2^2\right),
\ee
where  $+$ and $-$ denote the exterior and interior patches, respectively. For a spherical bubble one can identify the interior and the exterior spheres with the bubble itself.  However, $t$ and $r$ coordinates are in general discontinuous over $S$ and they should be differentiated in both regions. The third coordinate on $S$  can be chosen as the proper time $\t$. The hypersurface will be completely specified when the trajectories $t_\pm(\t)$ and $r_\pm(\t)$ are solved. By spherical symmetry, the induced metric $h_{ij}$ on $S$, where $i,j$ indices refer to the coordinates $(\t,\th,\phi)$, can be written as
\be
dh^2=-d\t^2+R(\t)^2\,d\O_2^2.\label{hm}
\ee
In the thin-shell approximation $h_{ij}$ is assumed to be continuous over $S$. Note that $R(\t)$ is the proper radius of the bubble at time $\t$. If one is only interested in the intrinsic properties, it would be enough to determine $R(\t)$. However, we are also concerned with the embedding of the bubble into the ambient spacetime.

By Einstein's field equations the discontinuity of the extrinsic curvature can be related to the energy momentum tensor on the hypersurface $S_{ij}$. Moreover, the four dimensional energy-momentum conservation also implies a conservation equation for  $S_{ij}$. One can show that these respectively yield the following equations for the bubble (known as junction conditions),
\bea
\left[-K_{ij}+Kh_{ij}\right]^{+}_{-}=S_{ij},\label{sfs}\\
D_iS^{ij}=\left[-T^{j}{}_{\m}n^\m\right]^{+}_{-},\label{scons}
\eea
where $D_i$ is the covariant derivative of $h_{ij}$.

In either region, the tangent vector $v^\m$ corresponding to the derivative operator $\del_\t$ can be calculated as
\be
v=t'\del_t+r'\del_r,
\ee
where prime denotes differentiation with respect to the proper time $\t$. By definition $v^\m v_\m=-1$. The two tangent vectors on the sphere together with $v^\m$ form a bases in the tangent space of $S$.
This information is enough to determine the unit normal vector $n^\m$ as
\be
n=\left(a\,r'\del_t+a^{-1}\,t'\del_r\right)\,\textrm{sign}(t'),
\ee
where the sign function is inserted to make sure that $n^\m$ is pointing from inside out, i.e. in the growing $r$ direction. Since $S$ is now uniquely specified, it is straightforward to calculate the extrinsic curvature which has components
\bea
&&K_{\t\t}=\left(ar't''-at'r''+a^2\fr{da}{dt}r'^3-2\fr{da}{dt}t'^2r'\right)\textrm{sign}(t'),\nn\\
&&K_{\th\th}=\left(a^2r^2\fr{da}{dt}r'+art'\right)\textrm{sign}(t')\label{ext}\\
&&K_{\phi\phi}=K_{\th\th}/\sin^2\th,\nn
\eea
where the last relation follows from spherical symmetry.

For a spherical false vacuum bubble associated with a minimally coupled scalar field,  the energy momentum tensor on the hypersurface can be assumed to have the form \cite{ts3}
\be
S_{ij}=-\s h_{ij}.
\ee
Thus in our problem there are five dynamical fields of interest which are $t_\pm(\t)$, $r_\pm(\t)$ and $\s(\t)$. The fact that $\t$ is the proper time gives two equations
\be
t_\pm'^2-a_\pm^2r_\pm'^2=1.\label{e1}
\ee
The continuity of the induced metric yields
\be
R^2=a_+^2r_+^2=a_-^2r_-^2,\label{e2}
\ee
where $R$ is defined in \eq{hm}. The conservation of stress energy tensor \eq{scons} can be satisfied provided
\be
\s'=\left[a\,t'\,r'\,(\rho+P)\right]^+_-\label{e3}
\ee
Finally $(\th\th)$ component of \eq{sfs} implies
\be
\left[(a^2\,r^2\fr{da}{dt}\,r'+a\,r\,t')\right]^+_-=-\fr{\s}{2}a^2\,r^2,\label{e4}
\ee
where we assume $\textrm{sign}(t')>0$ in both regions. Equations \eq{e1}, \eq{e2}, \eq{e3} and \eq{e4}  supply enough information to determine the dynamics of the bubble. Let us point out that $(\t\t)$ component of \eq{sfs} gives a second order equation
\be
\left[ar't''-at'r''+a^2\fr{da}{dt}r'^3-2\fr{da}{dt}t'^2r'\right]^+_-=\fr{\s}{2},
\ee
which is satisfied identically provided the above five equations and the Einstein's equations in the exterior and interior regions hold.

Following \cite{kyl} it is possible to obtain a useful equation for $R$. After some simple algebraic manipulations one can show that $(\th\th)$ component of the  discontinuity equation \eq{sfs}  implies
\be\label{dis2}
K_{\th\th}^{+2}=\fr{1}{\s^2R^4}\left(K_{\th\th}^{-2}-K_{\th\th}^{+2}+\fr{R^4\s^2}{4}\right)^2.
\ee
From \eq{ext} it follows that $K^2_{\th\th}$ can be expressed in terms of $R$ as
\be
K_{\th\th}^2=R^2\left[1+R'^2-H^2R^2\right]\label{k2} .
\ee
Using  \eq{k2} in \eq{dis2}  yields
\be
R'^2+1=\fr{R^2}{\s^2}\left(H_+^2-H_-^2\right)^2+\fr{R^2}{2}\left(H_+^2+H_-^2\right)
+\fr{R^2\s^2}{16}, \label{R}
\ee
where $H_\pm=d\ln a_\pm/dt_\pm$ are the corresponding Hubble parameters. Although this  looks like a single equation for $R$, let us note that there is a dependence on the variable $\s$ and a hidden dependence on $t_+$ through $H_+$. Note that $H_- \equiv H$ is constant in the interior de Sitter space.

It turns out that it is difficult to solve the above field equations to obtain an explicit solution for the boundary. Since our aim in this paper is to examine how  perturbations propagate from one region into the other, rather than studying  the bubble dynamics, we make some simplified assumptions.

First note that for local inflation it is desirable to have  $H_-\gg H_+$. Therefore, in \eq{R} one can ignore the terms involving $H_+$. This is nearly equivalent to taking outside region to be the flat space, except $\s$ is still time dependent. In \eq{e3}, the interior contribution  to the right hand side vanishes since the cosmological constant obeys $\rho_-+P_-=0$. By field equations, the exterior contribution  $\rho_++P_+$ is proportional to $dH_+/dt_+=-2/t_+^2$. We assume that  inflation occurs at a sufficiently later time after  big bang (i.e. $t_+\gg1$) such that this term is very small and thus $\s'\sim 0$. This effectively means that the outside region can be taken as the flat space. Summarizing  we take
\be
a_-=e^{Ht_-},\hs{4}a_+=1,\hs{4}\s'=0.
\ee
These assumptions may not be suitable for a realistic scenario using inflating bubbles  in a cosmological background, but they are not harmful for our purposes.

After these simplifications,  \eq{R} can be solved  for $R$ as
\be
R=\fr{1}{\a}\cosh(\a \t)
\ee
where
\be
\a=|\fr{H^2}{\s}+\fr{\s}{4}|
\ee
Since $a_+=1$, \eq{e2} and \eq{e1} can be used to determine $r_+$ and $t_+$ as
\be\label{asol1}
r_+=\fr{1}{\a}\cosh(\a \t),\hs{4} t_+=\fr{1}{\a}\sinh(\a \t).
\ee
To solve the interior fields we first note that \eq{e2} can be used to fix  $t_-$ in terms of $r_-$ as
\be\label{t-}
Ht_-=\ln R-\ln r_-.
\ee
Using \eq{t-} in the proper time equation \eq{e1}, one gets a quadratic equation for $\ln r_-'$ which can be solved algebraically as
\be\label{dif}
\ln r_-'=\fr{\ln R'\pm HR |\fr{H^2}{\s}-\fr{\s}{4}|}{1-H^2R^2}.
\ee
Although it looks complicated, this equation can be integrated to obtain  $r_-(\t)$. Further using \eq{t-} the interior embedding  coordinates can be fixed as
\bea
&&r_-=\fr{r_0\cosh(\a\t)}{|\sinh(\a\t)\pm\fr{4H^2-\s^2}{4H\s}|},\label{asol2}\\
&&Ht_-=\ln(|\sinh(\a\t)\pm\fr{4H^2-\s^2}{4H\s}|)-\ln(\a r_0).  \nn
\eea
The correlation between $\pm$ signs in \eq{asol2}  and \eq{dif}  depends on the sign of $4H^2-\s^2$. It may first be seen surprising that there appears two solutions since the initial system of differential equations are first order. However, it is easy to see that $\pm$  solutions are related by time reversal $\t\to-\t$. Without loss of any generality we choose $+$ sign in \eq{sol2}.

At this point we recall that although \eq{R} was obtained from the original junction condition, the two equations are not equivalent. Therefore, a final check is required to see whether  \eq{asol1} and \eq{asol2} obey \eq{e4}. It turns out that to satisfy \eq{e4} one should either choose $\s<0$ or interchange the interior and the exterior regions. Instead of choosing $\s<0$, which amounts to assume a negative cosmological constant on the shell, we simply switch the roles played by  the interior and exterior spaces.

\acknowledgments{The work of A.K. is partially supported by Turkish Academy of Sciences via Young Investigator Award Program (T\"{U}BA-GEB\.{I}P).}

\end{document}